\documentclass[conference]{IEEEtran}
\ifCLASSINFOpdf
  \usepackage[pdftex]{graphicx}
\else
\fi
\usepackage[cmex10]{amsmath}
\usepackage{mdwmath}
\usepackage{mdwtab}

\newcounter{MYtempeqncnt}

\begin{document}
%
\title{\LARGE A Geometrical Description of the SINR Region of the Gaussian Interference Channel: the two and three-user case}

\author{\IEEEauthorblockN{Abdoulaye Bagayoko and Patrick Tortelier}
\IEEEauthorblockA{Orange Labs (France Telecom R\&D)\\
38-40 Rue du General Leclerc, 92794 Issy-les-Moulineaux Cedex 9, France\\
Email:\{abdoulaye.bagayoko, patrick.tortelier\}@orange-ftgroup.com}
}


%


\maketitle
\begin{abstract}
This paper addresses the problem of computing the achievable rates for two (and three) users sharing a same frequency band without coordination and thus interfering with each other. It is thus primarily related to the field of cognitive radio studies as we look for the achievable increase in the spectrum use efficiency. It is also strongly related to the long standing problem of the capacity region of a Gaussian interference channel (GIC) because of the assumption of no user coordination (and the underlying assumption that all signals and interferences are Gaussian).  We give a geometrical description of the SINR region for the two-user and three-user channels. This geometric approach provides a closed-form expression of the capacity region of the two-user interference channel and an insightful of known optimal power allocation scheme. 
\end{abstract}
\IEEEpeerreviewmaketitle
\section{Introduction}
Interference is a fundamental issue in wireless communication when multiple uncoordinated links share a common communication medium. This paper addresses the problem of computing the achievable rates for two (or three) users sharing a same frequency band without coordination and interfering with each other. It is thus primarily related to the field of cognitive radio studies as we look for the achievable increase in the spectrum use efficiency. It is also strongly related to the long standing problem of the capacity region of a Gaussian interference channel (GIC) because of the assumption of no user coordination (and the underlying assumption that all signals and interferences are Gaussian).
Both topics have received a lot of attention in the technical literature where the interference channel is generally addressed via information theoretic tools, see for instance \cite{Spectrum_breaking}\cite{kramer}\cite{sason} and references herein. To this respect, \cite{Spectrum_breaking} proposes a  definition of cognitive radio as wireless system that makes use of "any available side information about activity, channel conditions, codebooks or messages used by the other users with which it shares the spectrum". 
What is the best performance one can achieve without making any a priori assumption on how the common resource is shared? We shall not assume any cooperation between users; they are not able to decode messages from other users, with the consequence that we shall not use sophisticated techniques 
such that dirty paper coding, rate splitting \cite{han-kobayashi} and their associated bounds for the achievable rates of each user. 
Due to its apparent simplicity, the two-user Gaussian interference channel (GIC) was the first to be addressed by the technical literature. Despite some special cases, such as very strong, strong ICs and the trivial case when there is no interference, in general the characterization of its capacity region is said an open problem. The exact characterization of the capacity region of the IC has been derived in the strong interference regime in \cite{carleial}, \cite{kramer} where it is shown that each user can decode the information transmitted to the other user.
The best known achievable strategy is the Han-Kobayashi scheme \cite{han-kobayashi}, where each user splits the information into private and common parts. The common messages are decoded at both the receivers, thereby reducing the level of interference. 
With the assumption of non cooperating users with power constrained Gaussian signals, the available rate of each of them is given by the $\log_2(1+SINR)$  classical formula, where SINR is the signal to noise plus interference ratio at the receiver. The difficulty we face is that the various SINRs of all users are not independent; they are interrelated in a way involving the channel coefficients as will be seen in the next section. Nevertheless we can have some insight in the shape (the geometry) of the set of possible SINRs, at least for the two or three-user interference channel. We can make use of this geometry to derive some new results: the capacity region of 
the two-user interference channel, and the SINR region of the three-user channel. Moreover, the way we derive this last result is very general and it allows deriving the $n$-user SINR region provided we know the one corresponding to $(n-1)$ users.
This geometric approach provides a closed-form capacity bounds expression of the two-user Gaussian interference channel when interference is considered as noise, although this strategy is known to be suboptimal. 

The remainder of this paper is organized as follows: we derive the analytical expressions of the capacity bounds, for the two-user Gaussian interference channel, in the Section II. In Section III, we tackle the problem of finding the maximum of the sum rate and we derive two possible areas in the plan where the maximum sum rate point can be. The three-user Gaussian interference channel is considered in Section IV where we find the analytical expressions characterizing the SINR region. Finally, conclusions are given in Section V.
\section{THE TWO-USER GAUSSIAN INTERFERENCE CHANNEL: CAPACITY REGION}
We consider a Gaussian interference channel with two transmitters and two receivers as depicted in the Fig.\ref{twoUserGIC}:
\begin{figure}[hbt!]
\centering
\includegraphics[width= 5cm]{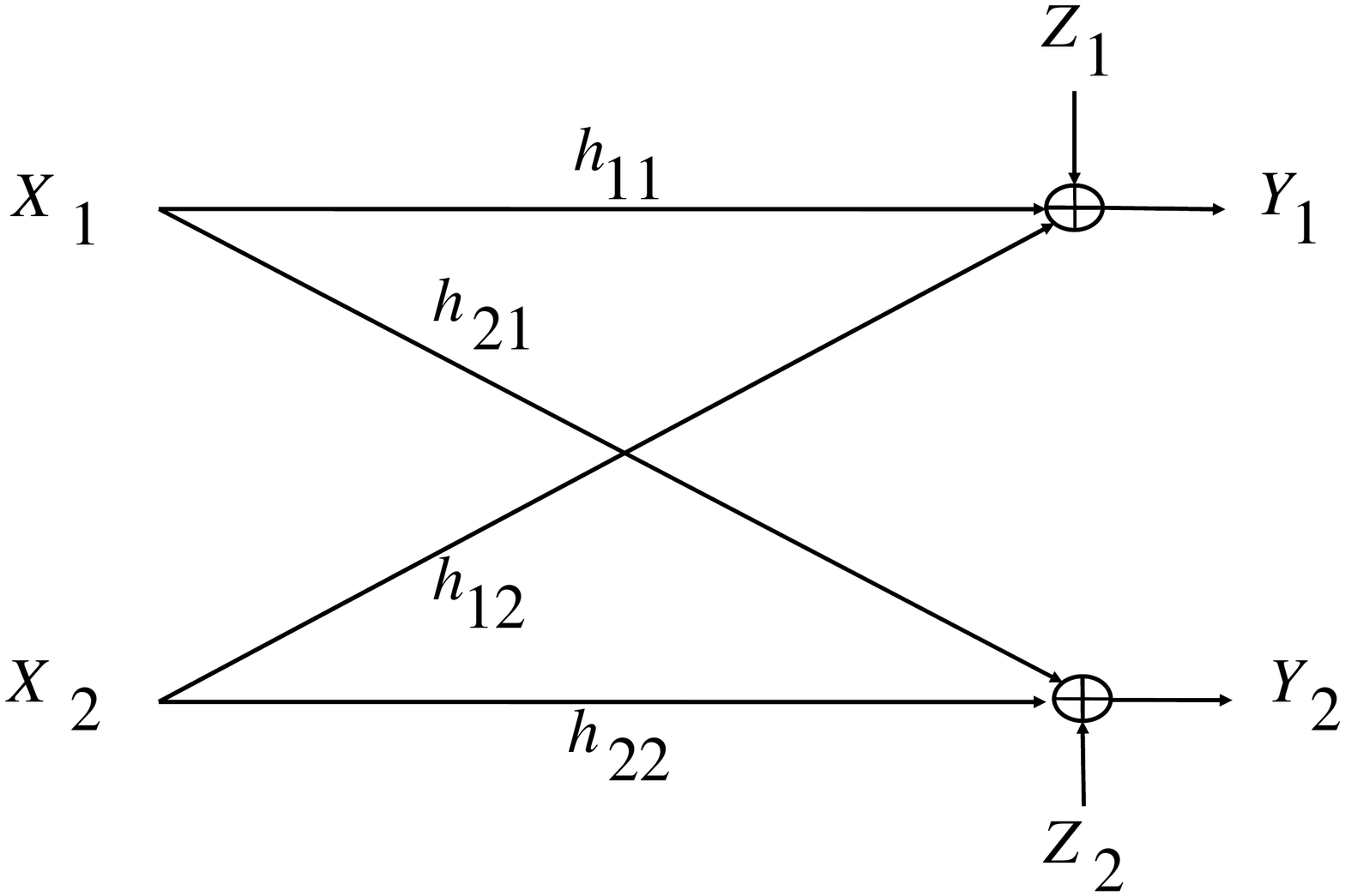}
\caption{The two-user Gaussian interference channel.}
\label{twoUserGIC}
\end{figure}
\begin{eqnarray}
Y_1 & = & h_{11}X_1 + h_{12}X_2 + Z_1\\
Y_2 & = & h_{21}X_1 + h_{22}X_2 + Z_2\nonumber
\end{eqnarray}
We shall assume that channel inputs are power-limited real Gaussian processes such that $p_i = E\left[ X_{i}^{2}\right]\leq P_i$ , and that there is no cooperation between users, so that interferences can be seen as Gaussian noise. With this assumption the two capacities of users 1 and 2 to their respective receivers are:
\begin{eqnarray}
C_1 &=& \frac{1}{2}\log_2\left(1 + \frac{g_{11}p_1}{\sigma^2 + g_{12}p_2} \right),\\
C_2 &=& \frac{1}{2}\log_2\left(1 + \frac{g_{22}p_2}{\sigma^2 + g_{21}p_1} \right),\hspace{0.5cm} g_{ij}=\left|h_{ij}\right|^2 \nonumber
\end{eqnarray}
With the change of variables $u_i = g_{ii}p_i/\sigma^2$ the above equations can be rewritten as:
\begin{eqnarray}
C_1 =\frac{1}{2}\log_2\left(1+S_1\right),& S_1=\frac{u_1}{1+a_{12}u_2}\\
C_2 =\frac{1}{2}\log_2\left(1+S_2\right),& S_2=\frac{u_2}{1+a_{21}u_1}\nonumber
\end{eqnarray}
where $a_{12}=g_{12}/g_{22}$ and $a_{21}=g_{21}/g_{11}$. The quantities $u_1$, $u_2$ are the SNR values when there is no interference and $S_1$, $S_2$ are the SINRs values (signal to noise plus interference ratios). The introduction of variables $u_1$, $u_2$ is similar to the introduction of the normalized channel in \cite{kramer} to which the reader is referred, as well as for an account of more results on the Gaussian interference channel.
The relation between the SINR variables $S_1$, $S_2$  and the SNR values $u_1$, $u_2$ can be easily inverted to obtain the two following expressions:
\begin{eqnarray}
u_1 &=& \frac{S_1(1+S_2a_{12})}{1-a_{12}a_{21}S_1S_2}\\
u_2 &=& \frac{S_2(1+S_1a_{21})}{1-a_{12}a_{21}S_1S_2}\nonumber
\end{eqnarray}
Expressing the power constraints $0\leq u_i \leq \bar{P_{i}}=g_{ii}P_i/\sigma^2$   allows us to derive corresponding constraints on the SINR variables, namely:
\begin{eqnarray}
\label{eq1} S_2 &\leq &\frac{1}{a_{12}a_{21}S_1} \\
\label{eq2} S_1 &\leq & \phi_1(S_2)=\frac{\bar{P_1}}{1+a_{12}S_2(1 + a_{21}\bar{P_1})} \\
\label{eq3}S_2 &\leq & \phi_2(S_1)=\frac{\bar{P_2}}{1+a_{21}S_1(1 + a_{12}\bar{P_2})}
\end{eqnarray}
The SINR region is thus delimited by these three curves. All variables being positive, the two functions $\phi_1(S_2)$  and $\phi_2(S_1)$ are respectively upper bounded by $(a_{12}\ a_{21}\ S_2)^{-1}$  and $(a_{12}\ a_{21}\ S_1)^{-1}$  so that the first inequality is redundant and is omitted in the sequel of the paper. The SINR region is then the intersection of the regions obeying each of the constraints defined by $\phi_1$, $\phi_2$ :
\begin{eqnarray}
\mathcal{D}'&= &\left\{(S_1, S_2)| 0\leq S_1 \leq \phi_1(S_2)\right\}\\
&&\cap \left\{(S_1, S_2)| 0\leq S_2 \leq \phi_2(S_1)\right\}\nonumber
\end{eqnarray}
We can also notice that $\phi_2(S_1)$ is simply obtained from $\phi_1(S_2)$ by the permutation $\{1,2\}\rightarrow \{2,1\}$, this result will be used later when considering the three-user case.
The second inequality (\ref{eq2}) above can be written in the equivalent form $S_2 \leq \left(\bar{P_1} -S_1 \right)/\left( a_{12}S_1\left(1+a_{21}\bar{P_1} \right) \right)$ so as to write the following analytic expression for the SINR region as a function of the sole $S_1$ :
\begin{equation}
0\leq S_2\leq \min\left(\frac{\bar{P_2}}{1 + a_{21}S_1\left(1+a_{12}\bar{P_2} \right)}, \frac{\bar{P_1}-S_1}{a_{12}S_1\left(1+a_{21}\bar{P_1}\right)}   \right)
\end{equation}
We shall use this expression to derive analytical bound to the capacity region of the interference channel.\\
The transformation $(u_1, u_2)\xrightarrow{\phi} (S_1,S_2)$  is a one to one correspondence of the region $\mathcal{D} =\{0\leq u_1 \leq \bar{P_1}, 0\leq u_2 \leq \bar{P_2} \}$  into the transformed region $\mathcal{D}'$, it leaves invariant the two points $(\bar{P_1},0)$ and $(0,\bar{P_2})$. We have $\mathcal{D}'\subset \mathcal{D}$, for $S_i \leq u_i$ . We can already notice that the more $\bar{P_1}$ and $\bar{P_2}$ increase the more the region $\mathcal{D}'$ will be constrained by the red curve in the Fig.\ref{SINRRegion} and its shape different from a rectangle.
\begin{figure}[hbt!]
\centering
\includegraphics[width= 8cm]{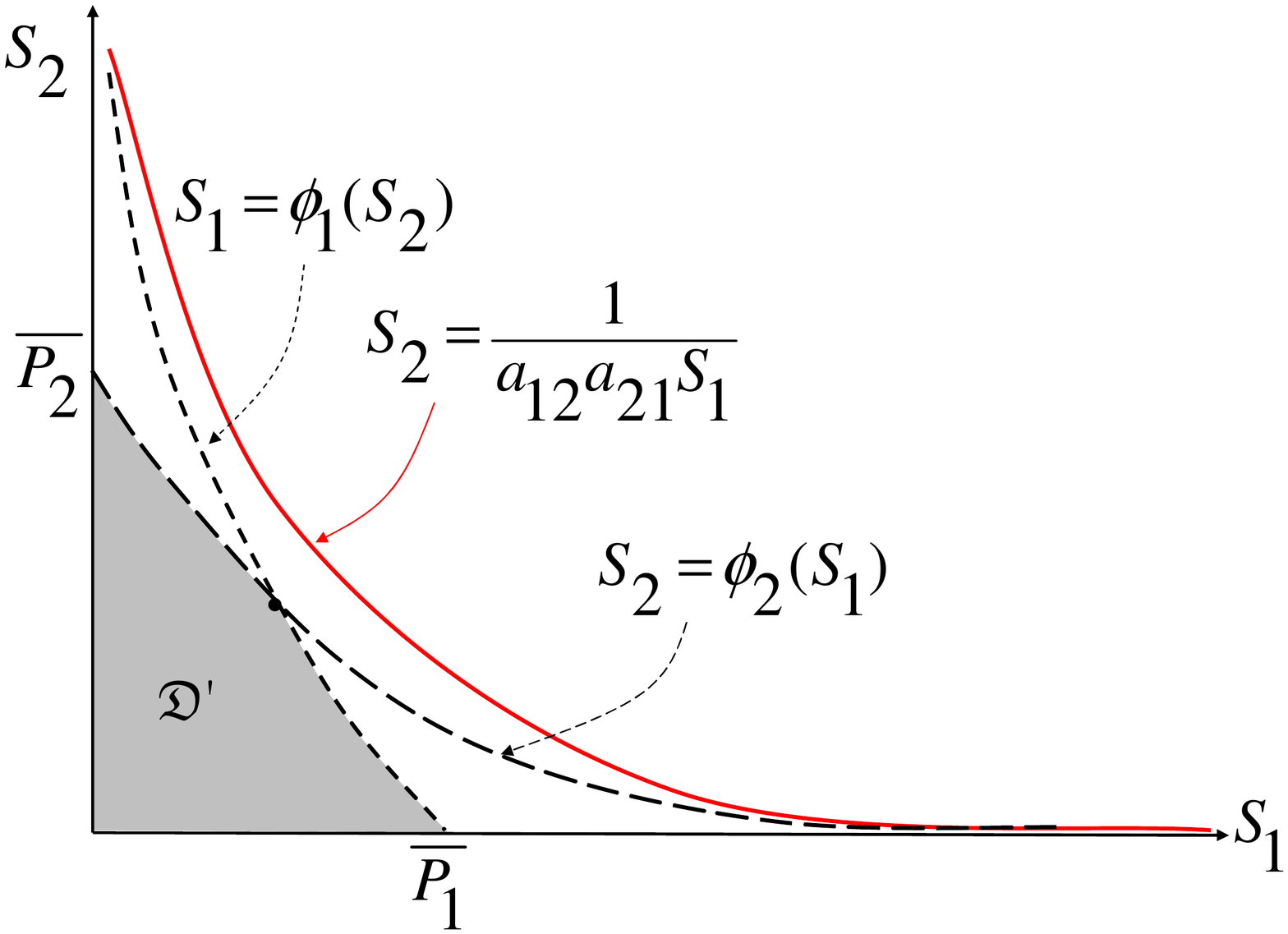}
\caption{Illustration of the SINR region for the two-user GIC.}
\label{SINRRegion}
\end{figure}
The last transform, $S_i\rightarrow \log_2(1+S_i)$, allows us to give an analytical expression of the capacity region boundary as a parametric curve rather than a simple function giving $R_2$ as a function of $R_1$  :
\begin{eqnarray}
0&\leq& t \leq \bar{P_1}\\
C_1 &=&\frac{1}{2}\log_2(1+t)\nonumber\\
C_2 &=&\frac{1}{2}\log_2(1+f(t))\nonumber\\
f(t) &=&\min\left(\frac{\bar{P_2}}{1+a_{21}t\left( 1 + a_{12}\bar{P_2}\right)}, \frac{\bar{P_1}-t}{a_{12}t\left( 1 + a_{21}\bar{P_1}\right)}\right)\nonumber
\end{eqnarray}
It is easy to check that $f(0)=\bar{P_2}$ and $f(\bar{P_1})=0$, that are the two cases where all capacity is allocated to only one user. As a result, the same parameterization provides an expression for the sum rate:
\begin{equation}
\mathbf{C_{SUM}}=\frac{1}{2}\log_2(1+t)+ \frac{1}{2}\log_2(1+f(t))
\end{equation}
Depending on the values of $\bar{P_1}$ and $\bar{P_1}$ and the coefficients of the normalized channel $a_{12}$, $a_{21}$, the capacity region and the sum capacity will exhibit different behaviors as depicted below for a symmetric case   $a_{12}=a_{21}$ (cf. Fig.\ref{caparegion1} and Fig.\ref{caparegion2}). The image of the point $(\bar{P_1},\bar{P_2})$ by the SNR to SINR transform is represented by a star; and the dashed curve is the constant sum rate line corresponding to the maximum $\mathbf{C_{SUM}}$ .

The following section is devoted to a more thorough analysis of these channel behaviors.
\begin{figure}[hbt!]
\centering
\includegraphics[width= 17cm]{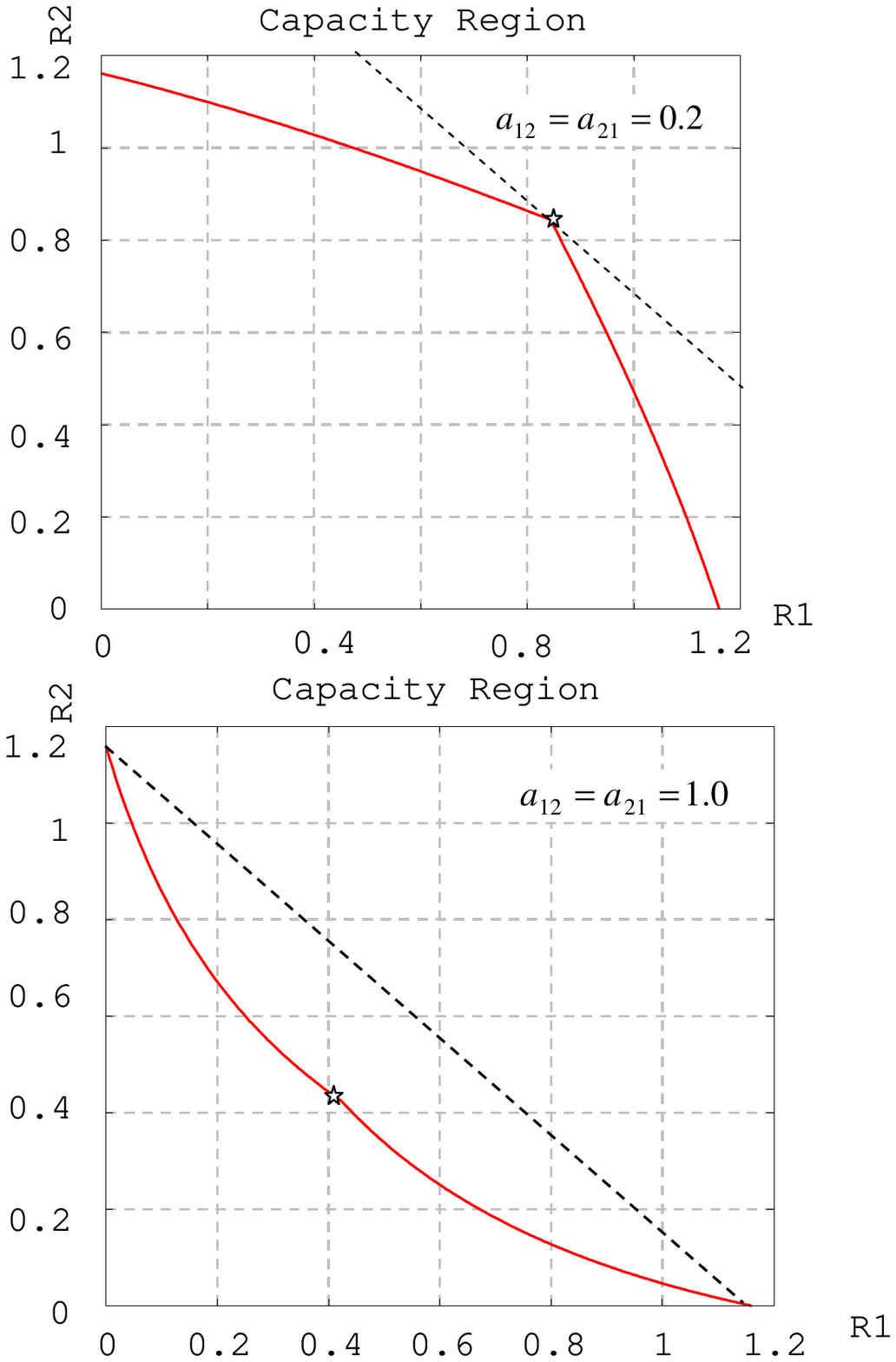}
\caption{The capacity region for $\bar{P_1}=\bar{P_2}=4$, medium interference (top) and strong interference (bottom).}
\label{caparegion1}
\end{figure}
\begin{figure}[hbt!]
\centering
\includegraphics[width= 17cm]{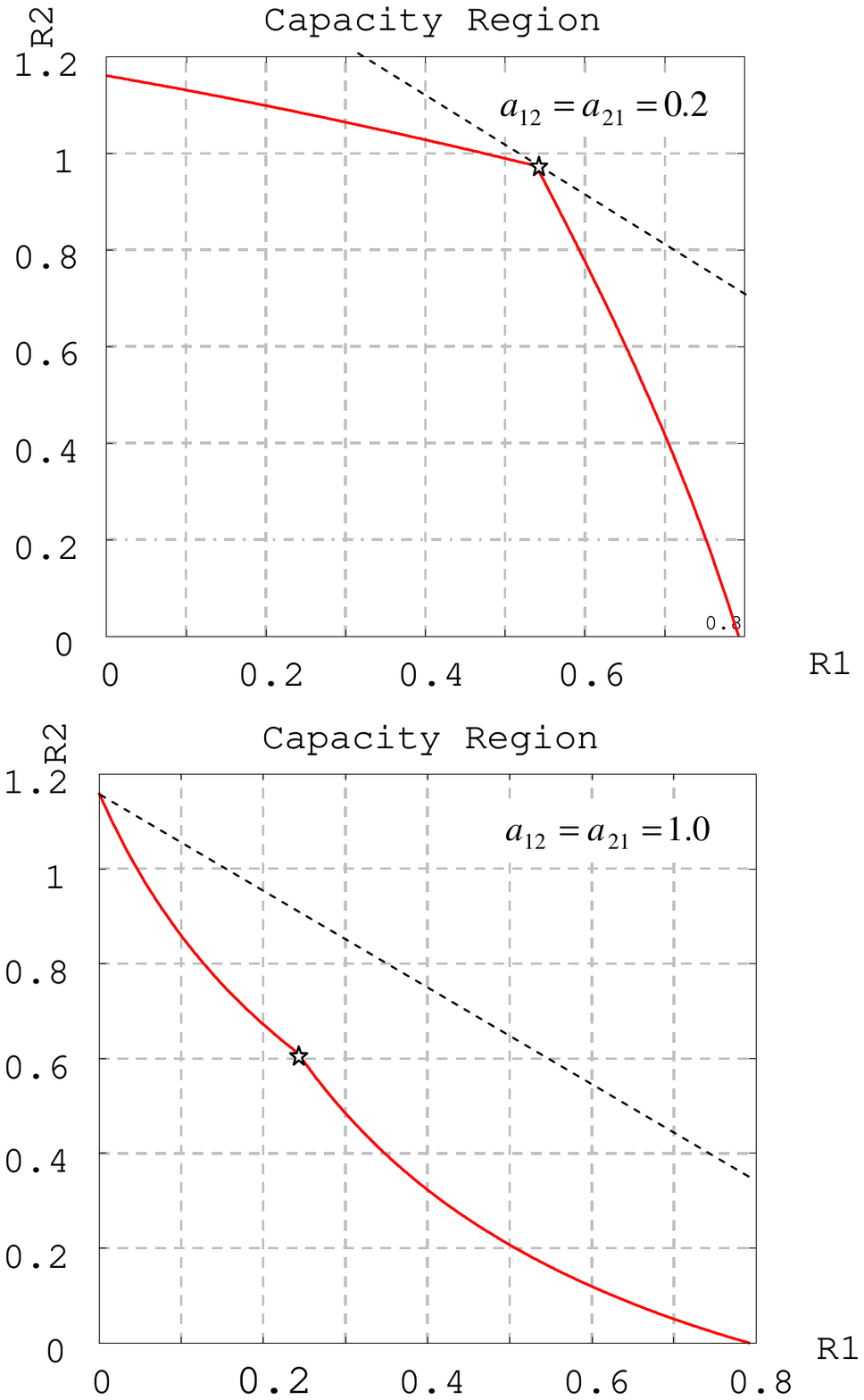}
\caption{The capacity region for $\bar{P_1}=2$,$\bar{P_2}=4$, medium interference (top), strong interference (bottom).}
\label{caparegion2}
\end{figure}
\section{SUM RATE MAXIMIZATION}
\begin{figure}[hbt!]
\centering
\includegraphics[width= 8cm]{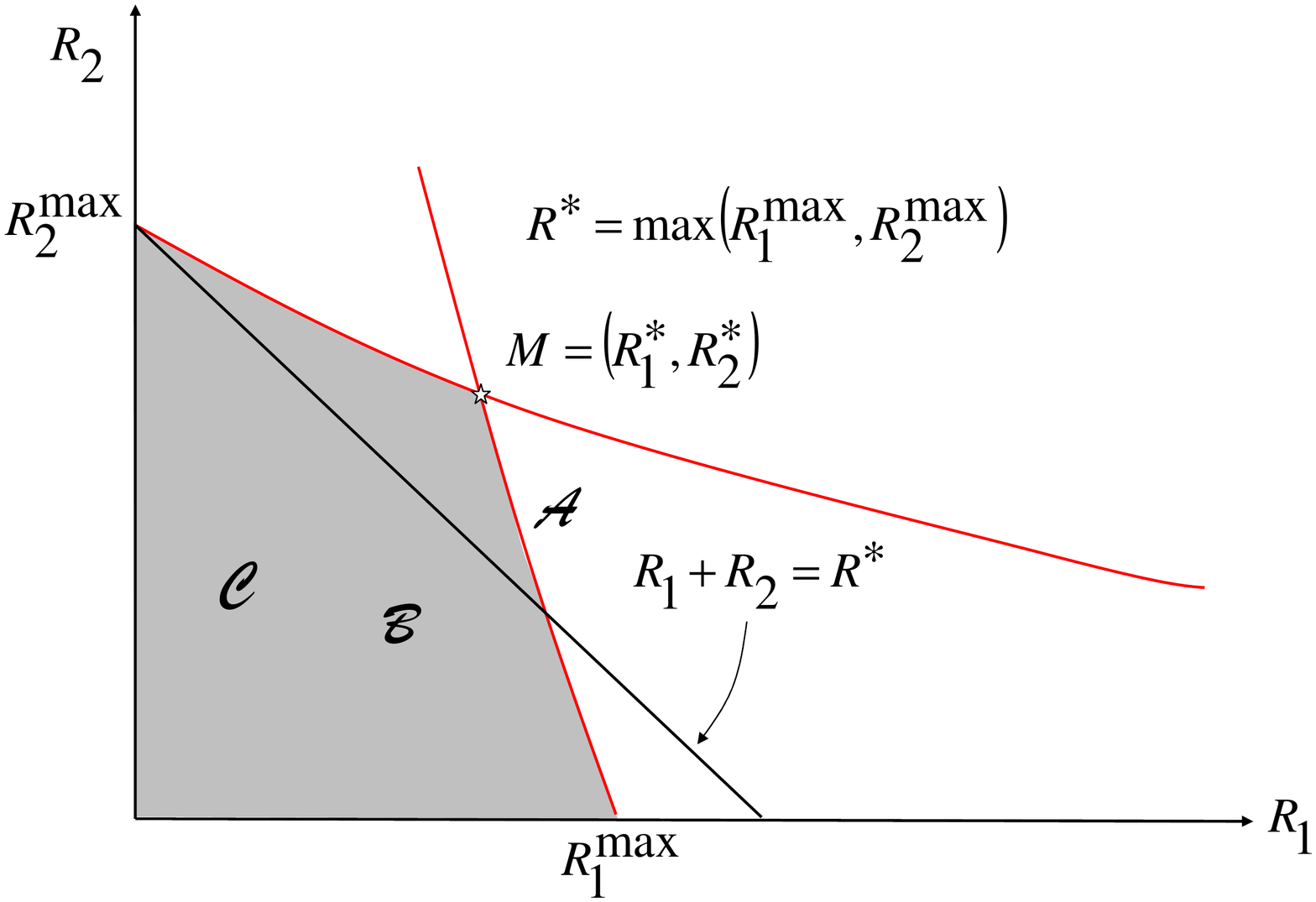}
\caption{The maximum Sum Capacity point for the two-user GIC.}
\label{sumcapa}
\end{figure}
In this part, we consider the maximization problem of the two-user sum rate expressed as a function of the two variables $u_1$, $u_2$ subject to the power constraints $u_{i}=\left(g_{ii}\ p_i/\sigma^2\right)\leq \bar{P_i} $:
\begin{eqnarray}
\mathbf{C_{SUM}} &=& C_1 + C_2\nonumber\\
&=&\frac{1}{2}\log_{2}\left(1 + \frac{u_1}{1+a_{12}u_2} \right)\nonumber\\
&& + \frac{1}{2}\log_{2}\left(1 + \frac{u_2}{1+a_{21}u_1} \right)
\end{eqnarray}
It is found in \cite{gesbert} that the optimal power allocation $(u^{*}_{1},u^{*}_{2})$ to this problem is one of the possible following vectors: $(0,\bar{P_{2}})$, $(\bar{P_{1}},0)$ or $(\bar{P_{1}},\bar{P_{2}})$. The same result is found in \cite{charafeddine} using the geometric programming method. 
Following our rate region analysis in section II, we derive two different regions $\mathcal{A}$ and $\mathcal{B}$ (cf. Fig. 5) such that:
\begin{enumerate}
    \item if the corner point $M\in \mathcal{A}$, then the optimal power allocation is $(\bar{P_1}, \bar{P_2})$;
    \item if $M\in \mathcal{B}$, then the optimal power allocation is $(\bar{P_1}, 0)$ or $(0, \bar{P_2})$.
\end{enumerate}
Denoting $R^* =\max\left( R_{1}^{max}, R_{2}^{max}\right)$, where $R_{i}^{max}=\frac{1}{2}\log_2\left(1+\bar{P_i} \right)$, the regions $\mathcal{A}$ and $\mathcal{B}$ are separated by the straight line with equation $R_1 + R_2 =R^*$. $\mathcal{A}$ is the region above the separator straight line and $\mathcal{B}$ is the region below.\\
Typically, since the point $M$ is reached for the rate vector $(R_{1}^{*}, R_{2}^{*})$, then\footnote{The rate vector $(R_{1}^{*}, R_{2}^{*})$ is reached when each user transmits with his maximum permitted power.} the maximum sum rate $R_{sum}^{max}$ verifies
\begin{equation}
R_{sum}^{max}\left\{\begin{array}{rlr}
=& R^*, & \mbox{if} (R_{1}^{*}+ R_{2}^{*})\leq R^*\\
=& R_{1}^{*} + R_{2}^{*}, & \mbox{if} (R_{1}^{*}+ R_{2}^{*})> R^*
\end{array}   
\right.
\end{equation}
The Fig.\ref{sumcapa} illustrates a case where the corner point $M \in \mathcal{A}$ and $(R_{1}^{*}+ R_{2}^{*})> R^*$, therefore the maximum sum rate $R_{sum}^{max}$ is reached for the power allocation $(\bar{P_1},\bar{P_2})$. 
\section{THE THREE-USER CASE}
When considering the three-user case, it is more convenient to write the relations between the SINR variables, $S_1$, $S_2$, $S_3$ and the SNR variables $u_1$, $u_2$, $u_3$ under the following form:
\begin{eqnarray}
u_1 &=& S_1\left(1+a_{12}u_2 + a_{13}u_3\right)\\
u_2 &=& S_2\left(1+a_{21}u_1 + a_{23}u_3\right)\\
u_3 &=& S_3\left(1+a_{31}u_1 + a_{32}u_2\right)
\end{eqnarray}
which we rewrite as a linear system of unknowns $(u_1, u_2, u_3)$:
\begin{equation}
\left(\begin{array}{ccc}
1 & -S_1a_{12} & -S_1a_{13}\\
-S_2a_{21} & 1 & -S_2a_{23}\\
-S_3a_{31} & -S_3a_{32}  & 1
\end{array} 
\right)\times \left(\begin{array}{c}
u_1\\
u_2\\
u_3
 \end{array} 
\right)=\left(\begin{array}{c}
S_1\\
S_2\\
S_3
 \end{array} 
\right)
\end{equation}
We can make use of the structure of the above $3\times 3$ matrix in order to make apparent the matrix $\mathbf{A}_2$ associated to the two-user problem:
\begin{equation}
\left\{ \begin{array}{l}
\mathbf{A}_3 =\left(\begin{array}{cc}
\mathbf{A}_2 & -\mathbf{a} \\
-S_3\mathbf{b}^t & 1
\end{array}\right)\\
\mathbf{A}_2 =\left(\begin{array}{cc}
1 & -S_1a_{12} \\
-S_2a_{21} & 1
\end{array}\right)\\
 \mathbf{a}=\left(\begin{array}{c}
S_1a_{13}\\
S_2a_{23}
\end{array}\right),\hspace{0.2cm} \mathbf{b}=\left(\begin{array}{c}
a_{31}\\
a_{32}
\end{array}\right)
\end{array}
\right.
\end{equation}
The linear system of unknowns $u_1$, $u_2$, $u_3$ can now be written as:
\begin{equation}
\left\{ \begin{array}{rll}
 \mathbf{A}_2\left(\begin{array}{c}
u_1\\
u_2
\end{array} 
\right)-\mathbf{a}u_3 &=& \left(\begin{array}{c}
S_1\\
S_2
\end{array} 
\right)\\
-S_3\mathbf{b}^t\left(\begin{array}{c}
u_1\\
u_2
\end{array} 
\right)+u_3 &=& S_3
\end{array}
\right.
\end{equation}
After some manipulations, and assuming that $\mathbf{A}_2$ is invertible we can express $u_3$ as:
\begin{equation}
u_3=S_3\times \frac{1+\mathbf{b}^t\mathbf{A}_{2}^{-1}\left(\begin{array}{c}
S_1\\
S_2
\end{array} 
\right)}{1-S_3\mathbf{b}^t\mathbf{A}_{2}^{-1}\mathbf{a}}
\end{equation}
From the constraint $u_3\leq \bar{P}_3$, we have, after some manipulations, a constraint on $S_3$ as a function of $S_1$ and $S_2$:
\begin{equation}
S_3 \leq \phi_3(S_1,S_2)=\frac{\bar{P_3}}{1 + (a_{31},a_{32})\mathbf{A}_{2}^{-1}\left(\begin{array}{c}
S_1(1+a_{13}\bar{P_3})\\
S_2(1+a_{23}\bar{P_3})
\end{array}    \right)}
\end{equation}
This is the equation of a surface in the three-dimensional space and it is worth noticing that when $S_1=0$ or $S_2=0$ the above upper bound becomes respectively equal to:
\begin{eqnarray}
S_3&\leq& \frac{\bar{P_3}}{1+a_{32}S_2(1+a_{23}\bar{P_3})}\\
S_3&\leq& \frac{\bar{P_3}}{1+a_{31}S_1(1+a_{13}\bar{P_3})}
\end{eqnarray}
In which we recognize the bounds already obtained for the two-user case when the two users are respectively $(2, 3)$ and $(1,3)$. A geometric representation of the constraints on $S_3$, when respectively $S_1=0$ and $S_2=0$, is shown in the Fig.\ref{georepconst}. As we also want to derive analogous relations for $S_1$  and $S_2$ we can make use of the invariance of the structure of the linear system under any permutation of the indexes $\{1, 2, 3\}$ to obtain the expressions (\ref{eqn_dbl_x}).
We shall denote these inequalities by $S_i\leq \phi_i(S_j, S_k)$ where $\{i, j, k\}$ is a permutation of the set $\{1, 2, 3\}$; with this notation the SINR region is the intersection of the three regions verifying respectively the three constraints:
\begin{eqnarray}
\mathcal{D}'&=&\mathcal{D}_1'\cap\mathcal{D}_2'\cap\mathcal{D}_3'\\
\mathcal{D}_1'&=&\left\{(S_1, S_2, S_3)| 0\leq S_1 \leq \phi_1(S_2, S_3) \right\}\nonumber\\
\mathcal{D}_2'&=&\left\{(S_1, S_2, S_3)| 0\leq S_2 \leq \phi_2(S_1, S_3) \right\}\nonumber\\
\mathcal{D}_3'&=&\left\{(S_1, S_2, S_3)| 0\leq S_3 \leq \phi_3(S_1, S_2) \right\}\nonumber
\end{eqnarray}
\begin{figure*}[hbt!]
\normalsize
\setcounter{MYtempeqncnt}{\value{equation}}
\setcounter{equation}{24}
\begin{eqnarray}\label{eqn_dbl_x}
S_3 &\leq & \bar{P_3}\frac{1-a_{12}a_{21}S_1S_2}{1-a_{12}a_{21}S_1S_2 + S_1(1 + a_{13}\bar{P_3})(a_{31} + S_{2}a_{32}a_{21}) + S_2(1+ a_{23}\bar{P_3})(a_{32} + S_1a_{31}a_{12})}\\
S_2 &\leq & \bar{P_2}\frac{1-a_{13}a_{31}S_1S_3}{1-a_{13}a_{31}S_1S_3 + S_1(1 + a_{12}\bar{P_2})(a_{21} + S_{3}a_{23}a_{31}) + S_3(1+ a_{32}\bar{P_2})(a_{23} + S_1a_{21}a_{13})}\nonumber\\
S_1 &\leq & \bar{P_1}\frac{1-a_{32}a_{23}S_3S_2}{1-a_{32}a_{23}S_3S_2 + S_3(1 + a_{31}\bar{P_1})(a_{13} + S_{2}a_{12}a_{23}) + S_2(1+ a_{21}\bar{P_1})(a_{12} + S_3a_{13}a_{32})}\nonumber
\end{eqnarray} 
\setcounter{equation}{\value{MYtempeqncnt}}
\hrulefill
\vspace*{4pt}
\end{figure*}
\begin{figure}[hbt!]
\centering
\includegraphics[width= 9cm]{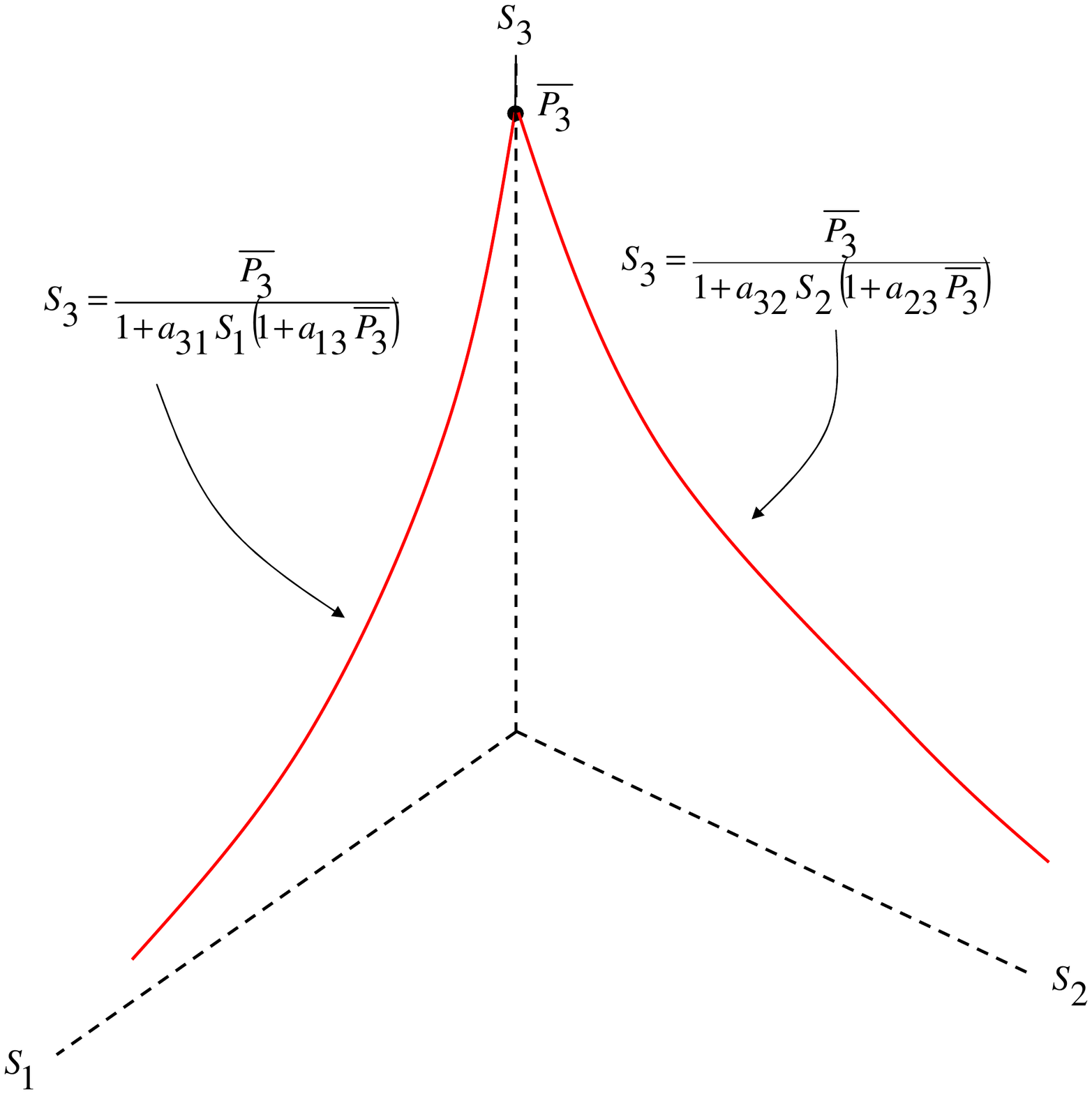}
\caption{A Geometric representation of the constraints on $S_3$, when respectively $S_1=0$ and $S_2=0$.}
\label{georepconst}
\end{figure}
In the Fig.\ref{illustsinr} we give a sketch of $\mathcal{D}'$ with the three sets of intersections on the faces of the positive quadrant.
\begin{figure}[hbt!]
\centering
\includegraphics[width= 9cm]{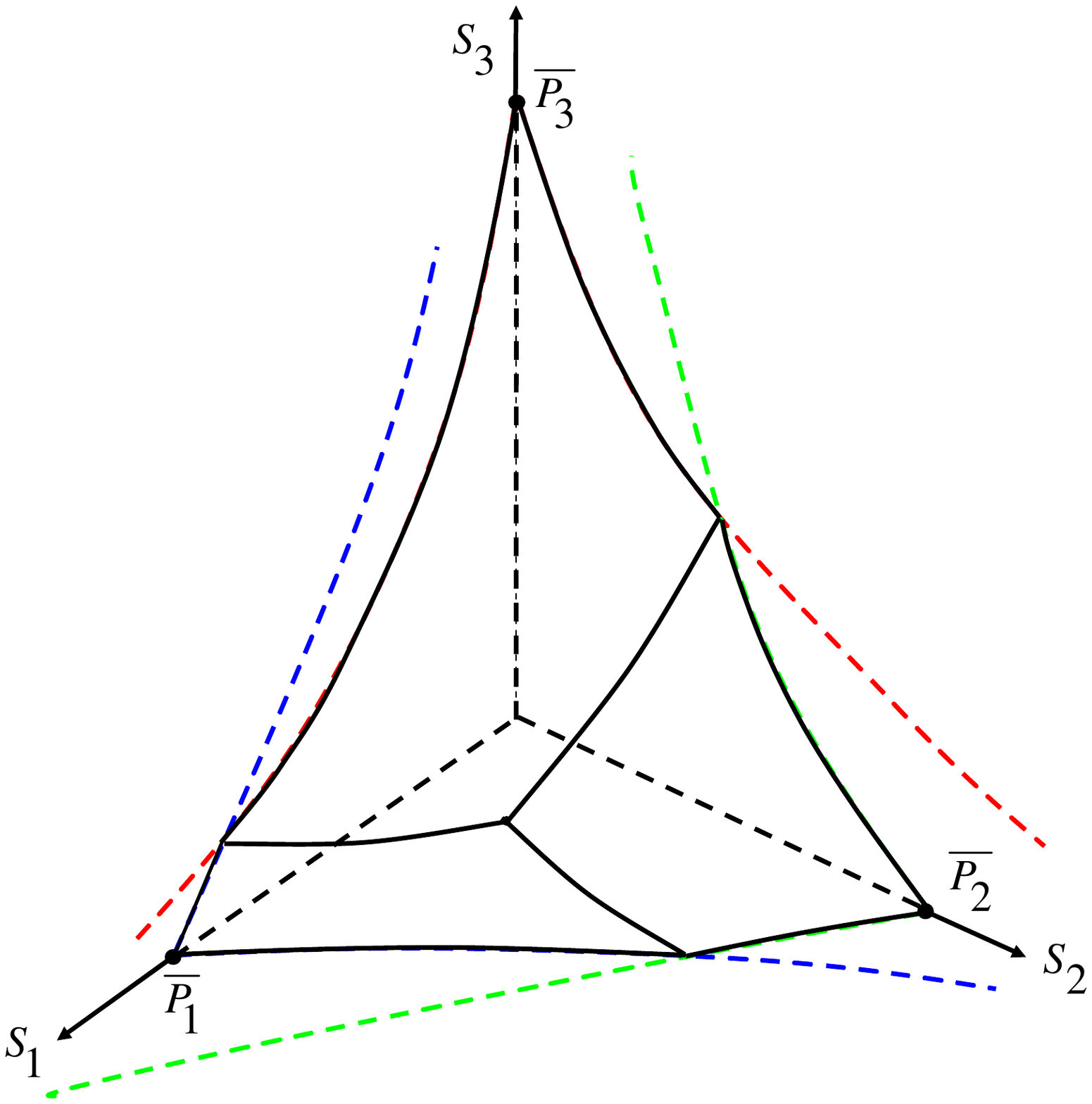}
\caption{Illustration of the SINR region for the three-user GIC.}
\label{illustsinr}
\end{figure}
\section{CONCLUSIONS}
\begin{figure}[hbt!]
\centering
\includegraphics[width= 10cm]{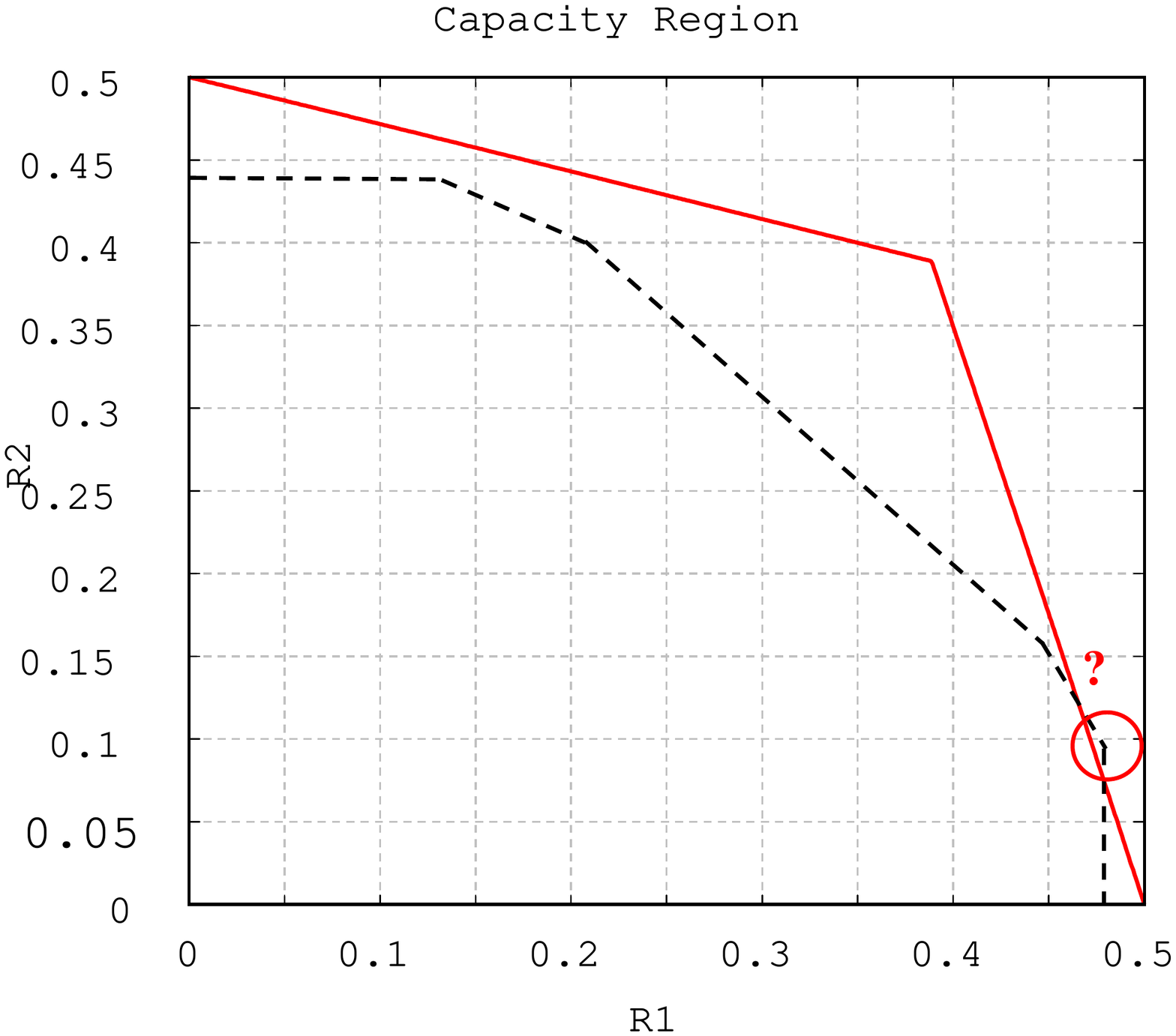}
\caption{Comparison of our result (red curve) with result from [12] (dashed curve: simplified HK bound) with parameters $a_{12}=a_{21}=0.4$ and $\bar{P_1}=\bar{P_2}=1$.}
\label{caparegionscomparison}
\end{figure}

In this paper, we derived the analytical expressions of the SINR bounds for the two and three-user Gaussian interference channel, treating the interference as noise. The way we derive the three-user SINR region is very general as it allows deriving the $n$-user case provided we know the result of the $(n-1)$ user case.  
Some examples show that an increase in the efficiency of the channel use is possible, depending upon the channel gains: the sum capacity of two-user is greater than the max capacity of a user alone, at the expense of a slight decrease of each user capacity. 
We have compared this solution to a modified Han-Kobayashi inner bound \cite{chong}; the comparison is given in the Fig.\ref{caparegionscomparison}. We see that, apart a dubious point due possibly to the limited accuracy of picking points in the original figure of \cite{chong} our capacity region contains the inner bound. Remains a question: our derivation of the capacity region does not involve any cooperation between the two users of the channel, we can expect that any techniques assuming partial knowledge of each user's message will improve the capacity region, that means it will contain our capacity region.


\begin{thebibliography}{1}
\bibitem{Spectrum_breaking}
A.~Goldsmith, S.~Jafar, I.~Maric and S.~Srinivasa.  Breaking spectrum gridlock with cognitive radios: an information theoretic perspective. \emph{Proceedings of the IEEE}, to appear 2008.
\bibitem{carleial}
A.B.~Carleial. A case where interference does not reduce capacity.\emph{IEEE Trans. Information Theory}, vol. IT-21, pp. 569-570, Sept. 1975.
\bibitem{kramer}
G.~Kramer. Review of rate regions for interference channels.\emph{Int. Zurich seminar on communications},pp. 162-165, Feb. 2006.
\bibitem{sason}
I.~ Sason. On achievable rate regions for the Gaussian interference channel.\emph{IEEE Trans. on Information Theory}, vol. 50, no. 6, pp. 1345 - 1356, June 2004.
\bibitem{han-kobayashi}
T.~Han and K.~Kobayashi. A new achievable rate region for the interference channel.\emph{IEEE Trans. Information Theory}, vol. 27, no.  1, pp. 49-60, Jan. 1981.
\bibitem{gesbert}
A.~Gjendemsjo, D.~Gesbert, G. E.~Oien and S. G.~Kiani.Optimal power allocation and scheduling for two-cell capacity maximization.\emph{IEEE WiOpt 2006}.
\bibitem{mahdavidoost}
H.~Mahdavidoost, M.~Ebrahimi and A.K.~Khandani. Characterization of Rate Region in Interference Channels with Constrained Power.\emph{Proc. IEEE International Symposium on Information Theory (ISIT'07)},pp. 2441-2445, Nice, France, June 24-29, 2007. 
\bibitem{motahariCapaBounds}
A. S.~Motahari and A. K.~Khandani.Capacity Bounds for the Gaussian Interference Channel.\emph{International Symposium on Information Theory (ISIT'08)}, Toronto, Canada.
\bibitem{shamai}
S.~Shamai and A. D.~Wyner.Information-theoretic considerations for symmetric, cellular, multiple-access fading channels.\emph{IEEE Trans. Information Theory}, vol. 43, N°6, Part I: pp. 1877-1894, Part II: pp. 1895-1911, Nov. 1997.
\bibitem{mathar}
L. A.~Imhof and R.~Mathar.The geometry of the capacity region for CDMA systems with general power constraints.\emph{IEEE Trans. Wireless Communications}, vol.4, N° 5,pp. 2040-2044, Sept. 2005.
\bibitem{charafeddine}
M.~Charafeddine and A.~Paulray.Sequential Geometric Programming for 2 X 2 Interference Channel Power Control.\emph{IEEE Information Science and Systems, CISS'07},41st Annual Conference,pp. 185-189, March 2007.
\bibitem{chong}
H. F.~Chong, H. K.~Garg and H.~El Gamal. On the Han-Kobayashi Region for the Interference Channel.\emph{IEEE Trans. On Information Theory}, vol. 54, no 7, pp. 3188-3195, July 2008.
\bibitem{sato}
H.~SATO.The capacity of the Gaussian interference channel under strong interference.\emph{IEEE Trans. Information Theory}, vol. IT-27, pp. 786-788,  Nov. 1981.
\end{thebibliography}
\end{document}